\documentclass[journal=nalefd,manuscript=letter,email=false]{achemso}

\usepackage{chemformula} 
\usepackage[T1]{fontenc} 
\usepackage{verbatim}

\usepackage{siunitx}               

	\title{All-optical noise spectroscopy of a solid-state spin}

\author{Demitry Farfurnik*}

\affiliation{Department of Electrical and Computer Engineering, Institute for Research in Electronics and Applied Physics, and Joint Quantum Institute, University of Maryland, College Park, MD 20742, USA}
\alsoaffiliation{Corresponding Author, dimka$\_$f13$@$yahoo.com}

\author{Harjot Singh}
\affiliation{Department of Electrical and Computer Engineering, Institute for Research in Electronics and Applied Physics, and Joint Quantum Institute, University of Maryland, College Park, MD 20742, USA}

\author{Zhouchen Luo}
\affiliation{Department of Electrical and Computer Engineering, Institute for Research in Electronics and Applied Physics, and Joint Quantum Institute, University of Maryland, College Park, MD 20742, USA}

\author{Allan S. Bracker}
\affiliation{Naval Research Laboratory, 4555 Overlook Avenue SW, Washington, D.C., 20375, USA.}

\author{Samuel G. Carter}
\affiliation{Naval Research Laboratory, 4555 Overlook Avenue SW, Washington, D.C., 20375, USA.}
\alsoaffiliation{Present address: Laboratory for Physical Sciences, University of Maryland, College Park, MD 20740, USA}

\author{Robert M. Pettit}
\affiliation{Department of Electrical and Computer Engineering, Institute for Research in Electronics and Applied Physics, and Joint Quantum Institute, University of Maryland, College Park, MD 20742, USA}
\alsoaffiliation{Intelligence Community Postrdoctoral Research Fellowship Program, University of Maryland, College Park, MD 20742, USA.}

\author{Edo Waks}
\affiliation{Department of Electrical and Computer Engineering, Institute for Research in Electronics and Applied Physics, and Joint Quantum Institute, University of Maryland, College Park, MD 20742, USA}


\begin{document}
\begin{abstract}
Noise spectroscopy elucidates the fundamental noise sources
in spin systems, thereby serving as an essential tool toward developing spin qubits with long coherence times for quantum information processing, communication, and sensing. But existing techniques for noise spectroscopy that rely on microwave fields become infeasible when the microwave power is too weak to generate Rabi rotations of the spin. Here, we demonstrate an alternative all-optical approach to performing noise spectroscopy. Our approach utilizes coherent Raman rotations of the spin state with controlled timing and phase to implement Carr-Purcell-Meiboom-Gill pulse sequences. Analyzing the spin dynamics under these sequences enables us to extract the noise spectrum of a dense ensemble of nuclear spins interacting with a single spin in a quantum dot, which has thus far only been modeled theoretically.  By providing spectral bandwidths of over 100 MHz, our approach enables the studies of spin dynamics and decoherence for a broad range of solid-state spin qubits.
\end{abstract}

	\maketitle
	\section*{}
   The coherent control of spin qubits can be used to conduct noise spectroscopy of their surrounding environment \cite{Bylander2011,Sung2021,BarGill2012,Romach2019,Malinowski2017,Chan2018}, in which the spin is used to probe the frequencies of the fluctuating fields generated by neighboring nuclear and electronic spins as well as the strength of their interaction with the spin (i.e., the spectral density).
Such noise spectroscopy using microwave fields has shed light on the fundamental noise processes of superconducting qubits \cite{Bylander2011,Sung2021}, nitrogen-vacancy centers in diamond \cite{BarGill2012,Romach2019}, and gate-defined quantum dots \cite{Malinowski2017,Chan2018}. The bandwidth of noise spectroscopy utilizing coherent control is limited by the rate of the spin rotation (i.e., the Rabi frequency), which must exceed the frequencies of the noise. While ultrafast Rabi spin rotations utilizing microwave control have been reported in nitrogen-vacancy centers \cite{Fuchs2009} and gate-defined  quantum dots \cite{Froning2021}, the realization of an efficient microwave control remains challenging for a broad range of solid-state spin systems that offer promising optical  properties for photonic quantum information processing and toward the realization of quantum networks. These systems include optically-active quantum dots in a variety of direct bandgap semiconductors \cite{Bayer2002,Huber2017,Benyoucef2013} and color center spins with large spin-orbit couplings in wide bandgap materials \cite{Debroux2021}. In the lack of an efficient microwave control, an alternative approach is required to characterize the noise sources of such promising spin qubits.

   An alternative approach for noise spectroscopy is to use optical fields. For example, optical measurements of resonance fluorescence \cite{Kuhlmann2013,Munsch2017} and Faraday rotation \cite{Dahbashi2014} have quantified the impact of noise on the dynamics of single spins without applying coherent spin control. However, these approaches do not operate within the ground state energy manifold of the spin qubit and do not quantify the strength of interactions  of different noise frequency components with the spin, as required for the full understanding of spin dynamics and decoherence. In addition, the bandwidth of noise spectroscopy techniques that measure resonance fluorescence or Faraday rotation on a single spin remained slower than the high frequencies (up to 100 MHz \cite{Bulutay2012,Stockill2016,Chekhovich2012,Bulutay2014}) associated with the noise sources that dictate spin dynamics and decoherence. Probing such high frequency noise of a single spin requires the application of sequences of all-optical coherent control pulses. 

     In this Letter, we demonstrate all-optical noise spectroscopy of a single solid-state spin utilizing coherent control. Our approach utilizes Raman rotation pulses with controllable timing, amplitude, and phase \cite{Bodey2019,Gangloff2019} to implement the Carr-Purcell-Meiboom-Gill (CPMG) control sequences \cite{Meiboom1958}. With this approach, we probe the noise source of an ensemble of indium and arsenic nuclear spins (i.e., the Overhauser field) interacting with a single electron spin confined in an InAs/GaAs quantum dot, to extract its spectral density thus far only modeled theoretically \cite{Bulutay2012,Stockill2016}. Previously demonstrated microwave control of InAs/GaAs quantum dots has been orders of magnitude slower than required for noise spectroscopy due to the low g-factor ($\sim$ 0.4) \cite{Kroner2008} of the electron confined in the quantum dot. In contrast, the high spectral bandwidths of noise spectroscopy ($>100$ MHz) provided by our all-optical approach enable us to probe the Overhauser field, which features high frequencies due to the spread of the Larmor frequencies of the background nuclei at high magnetic fields. The extracted spectra verify a theoretical model that predicts two noise components (in-plane and out-of-plane to the external field) caused by an inhomogeneous strain field in the quantum dot environment \cite{Bulutay2012,Stockill2016}. Understanding of such noise sources can quantify the achievable coherence times of spin qubits and predict the spin dynamics of such qubits for quantum information processing, communication, and sensing.  

\begin{figure}
	\centering
	\includegraphics[width=0.8\textwidth]{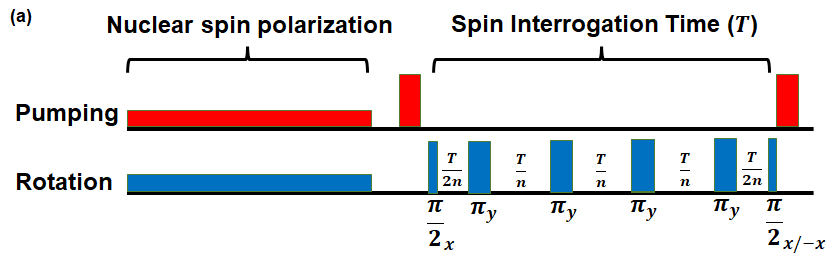}\\
	\includegraphics[width=0.35\textwidth]{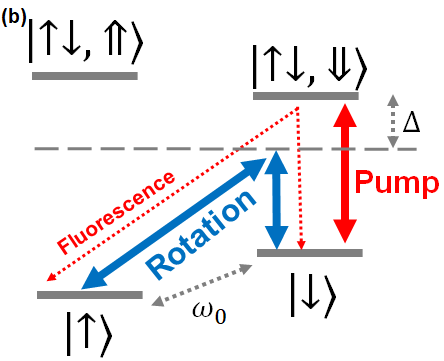}
	\includegraphics[width=0.40\textwidth]{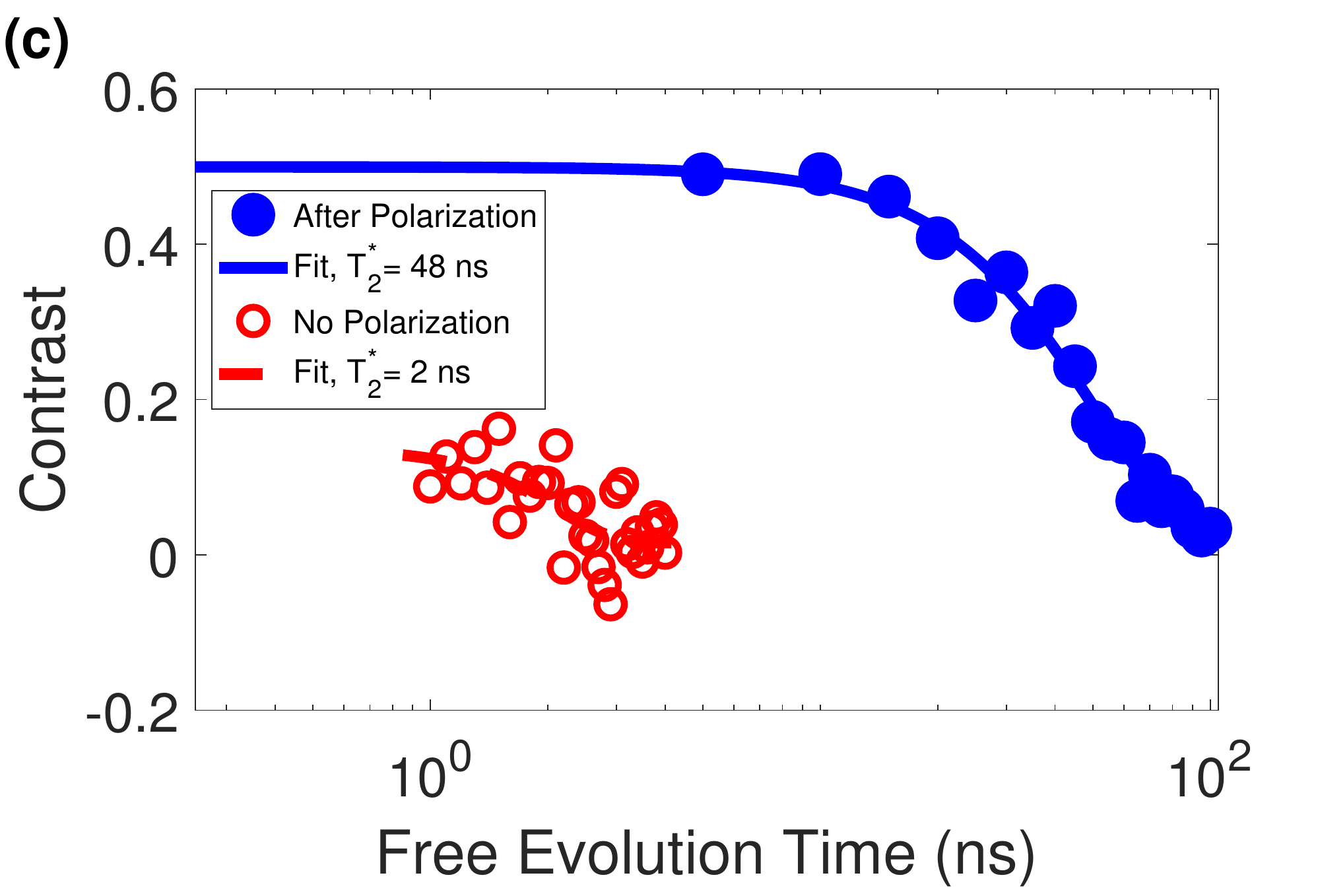}
	\caption{
The realization of the CPMG pulse sequences used for noise spectroscopy. (a) Representative scheme of a CPMG sequence, with $ n = 4$ pulses and total interrogation time $T$, performed after 4 \SI{}{\micro\second} of nuclear polarization. The phase of the last pulse is alternated between $0^o$ and $180^o$ for normalization purposes. The pumping and rotation amplitudes for nuclear polarization (illustrated by the heights of the pulses) are weaker than those used for spin interrogation. (b) The energy-level diagram of a negatively charged quantum dot under an external magnetic field in the Voigt geometry, including the optical pumping (solid red arrows) and Raman rotation (solid blue arrows) transitions used for noise spectroscopy, as well as the optical decay transitions from the excited state (dashed red arrows). Fluorescence collected from the   $\lvert \uparrow\downarrow,\Downarrow\rangle\rightarrow\lvert\uparrow\rangle$ optical decay transition indicates the quantum dot spin state. (c) The Ramsey         experiment used to characterize our preliminary polarization of the nuclear ensemble. The resulting spin dynamics without (dashed red line) and with (solid blue line) polarization of the nuclear spin environment at an external field of $B = 2.4$ T fitted to the function $a e^{-(T/T^*_2)^p}$ shows the electron spin coherence time can be extended from $T_2^*\approx 2$  ns (dashed red line) to $T_2^*\approx 48$ ns (solid blue line). From the fits, we extract $a\approx 0.13$ and $a\approx 0.5$ with and without the nuclear polarization, respectively, and $p\approx 1.85$ for both curves.}
	\label{fig:fig1}
\end{figure}

We perform the all-optical noise spectroscopy using a negatively charged quantum dot (Section I of the Supporting Information) \cite{Bayer2002,Gao2012,Ding2016,Somaschi2016,Lodahl2018,Scholl2019,Tomm2021,Appel2021}, which acts as a quantum probe for the fluctuations of nuclear spins in its environment. Under an external magnetic field applied perpendicular to the sample growth direction (Voigt geometry), the electronic structure of these dots consists of an electron ground-state spin qubit $(\{\lvert\uparrow\rangle,\lvert\downarrow\rangle\})$ and two optically excited trion states $(\{\lvert\uparrow\downarrow,\Uparrow\rangle,\lvert\uparrow\downarrow,\Downarrow\rangle\})$ \cite{Bayer2002}. Spontaneous decay from the excited states leads to fluorescence emission of single photons with high efficiency and indistinguishability \cite{Ding2016,Somaschi2016,Scholl2019,Tomm2021}. The main noise source that leads to the decoherence of the quantum dot spin, $S(\omega)$,  is the Overhauser field of indium and arsenic nuclear spins, consisting of frequencies of $\sim$ tens of MHz \cite{Bulutay2012,Stockill2016} (Section II of the Supporting Information). Here, we perform noise spectroscopy of the Overhauser field utilizing optical CPMG pulse sequences.          

 The CPMG sequences \cite{Meiboom1958} are illustrated in Figure 1a. After a preliminary 4-\SI{}{\micro\second} long stage of nuclear spin polarization \cite{Gangloff2019}, we initialize the spin in the $\frac{\lvert\uparrow\rangle+\lvert\downarrow\rangle}{\sqrt{2}}$ state using an optical pumping pulse followed by a $\frac{\pi}{2}$-rotation pulse (Section I of the Supporting Information). Then, we apply $n\geq 1$ equally spaced $\pi$-pulses that modify the temporal dynamics of the spin. After the application of a second $\frac{\pi}{2}$-pulse, a final optical pumping pulse induces a fluorescence signal that indicates the final spin state. We define the coherence function, $C(T)$, as the decay of the fluorescence signal as a function of the spin interrogation time, $T$ (i.e., the time between the $\frac{\pi}{2}$-pulses). 

   To implement the all-optical coherent control pulses required for the CPMG sequences, we utilize detuned two-photon Raman excitations \cite{Bodey2019} (solid blue arrows in Figure 1b) enabled by a modulated laser. We perform this modulation using an arbitrary waveform generator, which introduces up to eight spin rotation $\pi$-pulses with precise timing, phase, and Rabi frequencies of up to $\approx$ 150 MHz (see Section III of the Supporting Information for the optimization of the parameters). Leveraging the arbitrary modulation capabilities with sampling rates of up to 65 Gs/s allows us to realize CPMG sequences at short times and desired intervals, which is essential for obtaining noise spectra with sufficiently high bandwidth and spectral resolution (Section IV of the Supporting Information). 
   
   To ensure that spin dephasing is minimal during the application of the pulses, we use the Raman coherent control to polarize the nuclear spin ensemble prior to the application of any CPMG sequence \cite{Gangloff2019}. This step suppresses the uncertainty in the total magnitude of the Overhauser field at the beginning of an experiment, thereby reducing the inhomogeneous dephasing of the quantum dot spin. As shown by a Ramsey experiment (Figure 1c; see Section I of the Supporting Information), such a 4-\SI{}{\micro\second} long nuclear polarization step increases the inhomogeneous dephasing time of the quantum dot spin from $T_2^*\approx 2$  ns (dashed red line in Figure 1c) to $T_2^*\approx 48$ ns (solid blue line in Figure 1c), consistent with previous Ramsey experiments \cite{Gangloff2019}. After completing the nuclear polarization step, the inhomogeneous dephasing time is an order of magnitude longer than the spin rotation $\pi$-pulses, which allows us to use such pulses for the realization of CPMG sequences for high-frequency noise spectroscopy.

 We first apply the simplest form of the CPMG sequence, namely the Hahn-echo experiment consisting of a single $\pi$-pulse (Figure 2a). Consistent with previous Hahn-echo measurements on quantum dots \cite{Stockill2016,Press2010,Bechtold2015}, the decay timescale of the spin dynamics (i.e., the coherence time) increases as a function of the external magnetic field, B, up to $T_2\approx 1$ \SI{}{\micro\second} (blue dots in Figure 2a). This increase is associated with the Zeeman terms of the indium and arsenic nuclei dominating over the inhomogeneous broadening of these nuclei at high magnetic fields (Section II of the Supporting Information). Furthermore, the spin exhibits a two stage decay in its coherence that consists of a fast drop of the signal contrast (at $T\approx 30$ ns), followed by a second decay (starting at $T\approx 100$ ns). This behavior is analogous to previously observed Hahn-echo spin dynamics of quantum dots \cite{Stockill2016,Bechtold2015}. Intuitively, the two stages of decoherence suggest that separate spectral components of the noise affect the spin dynamics at separate timescales. Simulation results (the lines in Figure 2a), which consider such separate noise components associated with the strained nuclear environment of the quantum dot \cite{Bulutay2012,Stockill2016} (Section II of the Supporting Information), agree with the experimental results, thereby confirming this hypothesis. To  experimentally extract these spectral noise components, we apply CPMG sequences with increasing numbers of pulses.

\begin{figure}
	\centering
	\includegraphics[width=0.4\textwidth]{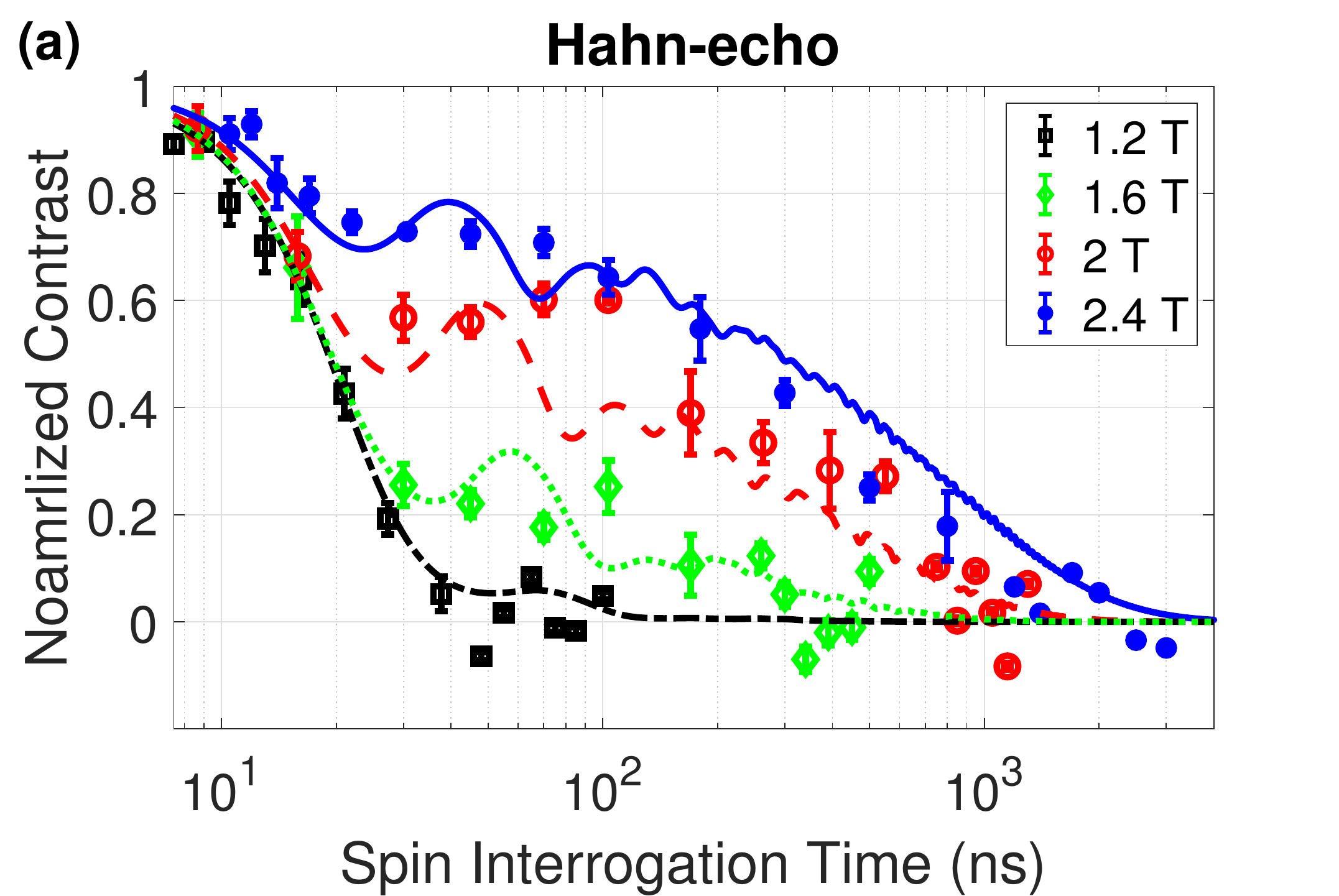}
	\includegraphics[width=0.4\textwidth]{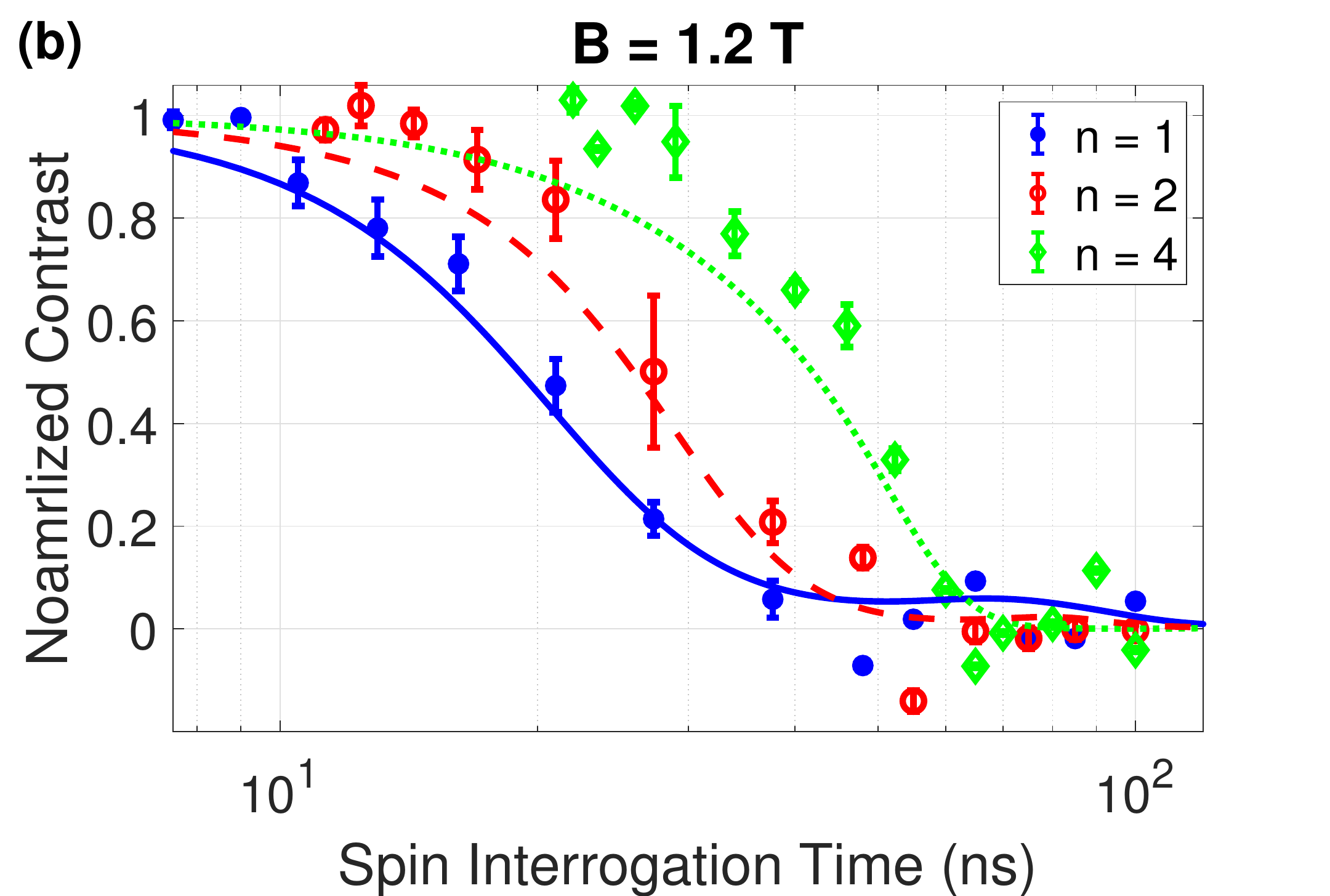}\\
	\includegraphics[width=0.4\textwidth]{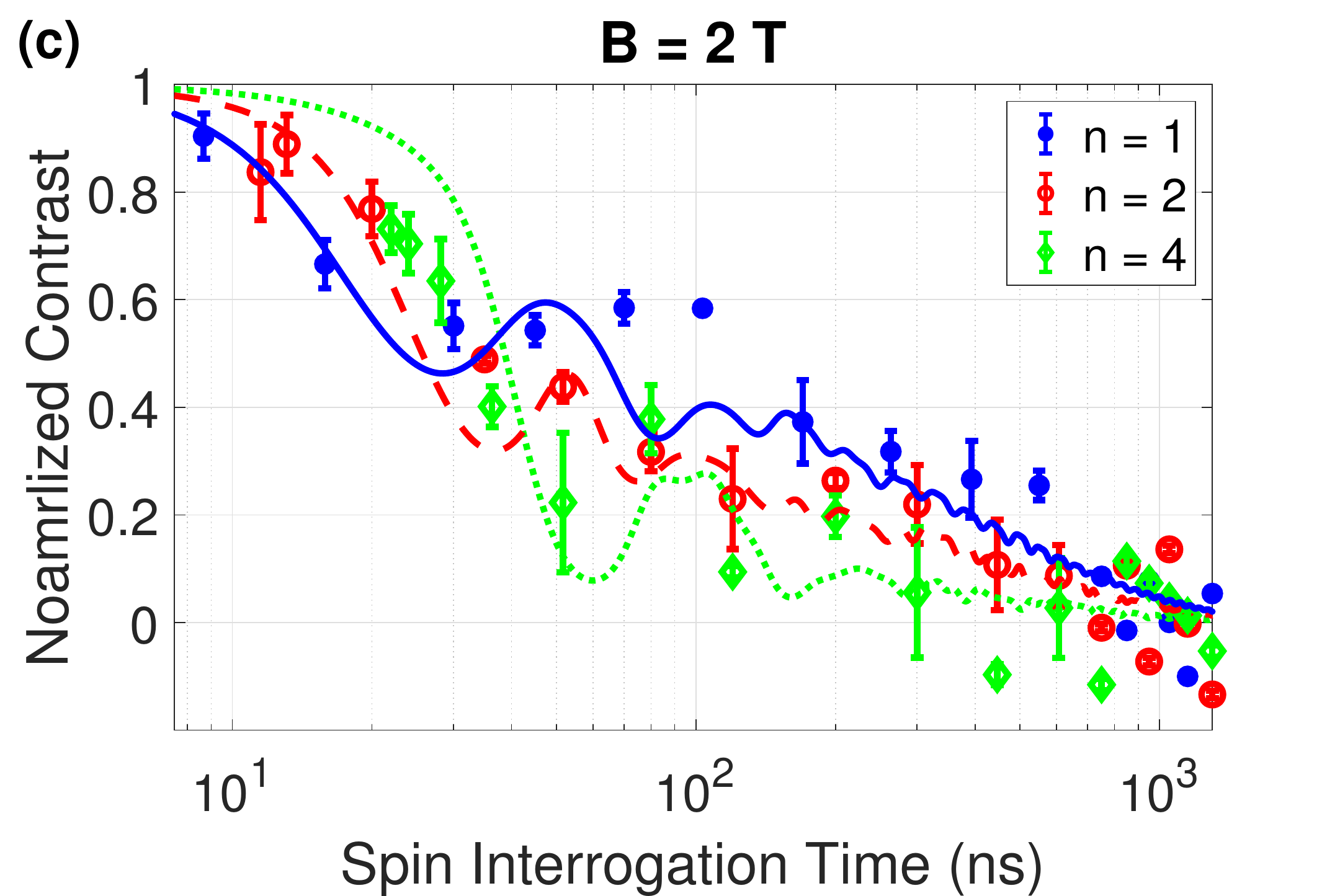}
\caption{
		Normalized coherence functions of a quantum dot spin under the application of CPMG sequences with $n$ pulses. (a) The  results of the Hahn-echo sequence $(n = 1)$ at different magnetic fields, $B$. (b)-(c) The results of multi-pulse CPMG sequences at $B = 1.2$ T (b) and $B = 2$ T (c). In all sub figures, the data points represent experimental data and the lines represent theoretical simulations. The error bars in the experimental data originate from their (x2) binning for the clarity of their presentation.
}
	\label{fig:fig2}
\end{figure}

The temporal dynamics of the quantum dot spin under the application of such sequences are presented in Figure 2b and Figure 2c for the external magnetic fields of 1.2 T and 2 T, respectively. As shown in Figure 2b for $B = 1.2$ T, the measured spin coherence times increase with $n$ as the decay timescale of the coherence function becomes longer. However, as shown in Figure 2c, the two-stage spin decoherence profile for $B = 2$ T exhibits a more complicated behavior as a function of $n$. Our simulations (the lines in Figure 2b-c) considering two spectral components of the Overhauser field \cite{Bulutay2012,Stockill2016} agree with the observed experimental behavior. However, to understand these complex dynamics requires a comparison between the rate of application of these pulses with the frequencies of the noise spectra \cite{Bylander2011,BarGill2012,Romach2019,Malinowski2017,Chan2018,Stockill2016,Farfurnik2015,Cywinski2008,deSousa2009}.   

 We experimentally extract these noise frequencies by analyzing the all data collected from four different measurements of the coherence functions under the application of CPMG sequences with $n = 1, 2, 4$ and $8$ $\pi$-pulses  utilizing a recursive numerical integration method (Section IV of the Supporting Information). The extracted spectral densities, plotted for external magnetic fields between 1.2 T and 2 T (blue dots in Figure 3a-c), display a broad range of frequencies of up to 100 MHz. To characterize the behavior of the noise as a function of the magnetic field, we fit the extracted spectra to Gaussian functions (solid blue lines in Figure 3a-c). The central frequency of the noise (center of the Gaussian fit) increases with the magnetic field up to 38 MHz at $B = 2$ T (Figure 3d), corresponding to (twice) the Larmor frequency of indium nuclei \cite{Chekhovich2012} (presented in Table 1 of the Supporting Information) that dominate the interaction with the electron spin due to their spin-9/2 nature. Meanwhile, the amplitude of the noise (at the central frequency) decreases with the magnetic field (Figure 3e) as nuclear Zeeman interactions dominate over the broadening of the nuclei due to strain fields (Section II of the Supporting Information) \cite{Bulutay2012,Stockill2016}.

\begin{figure}
	\centering
	\includegraphics[width=0.32\textwidth]{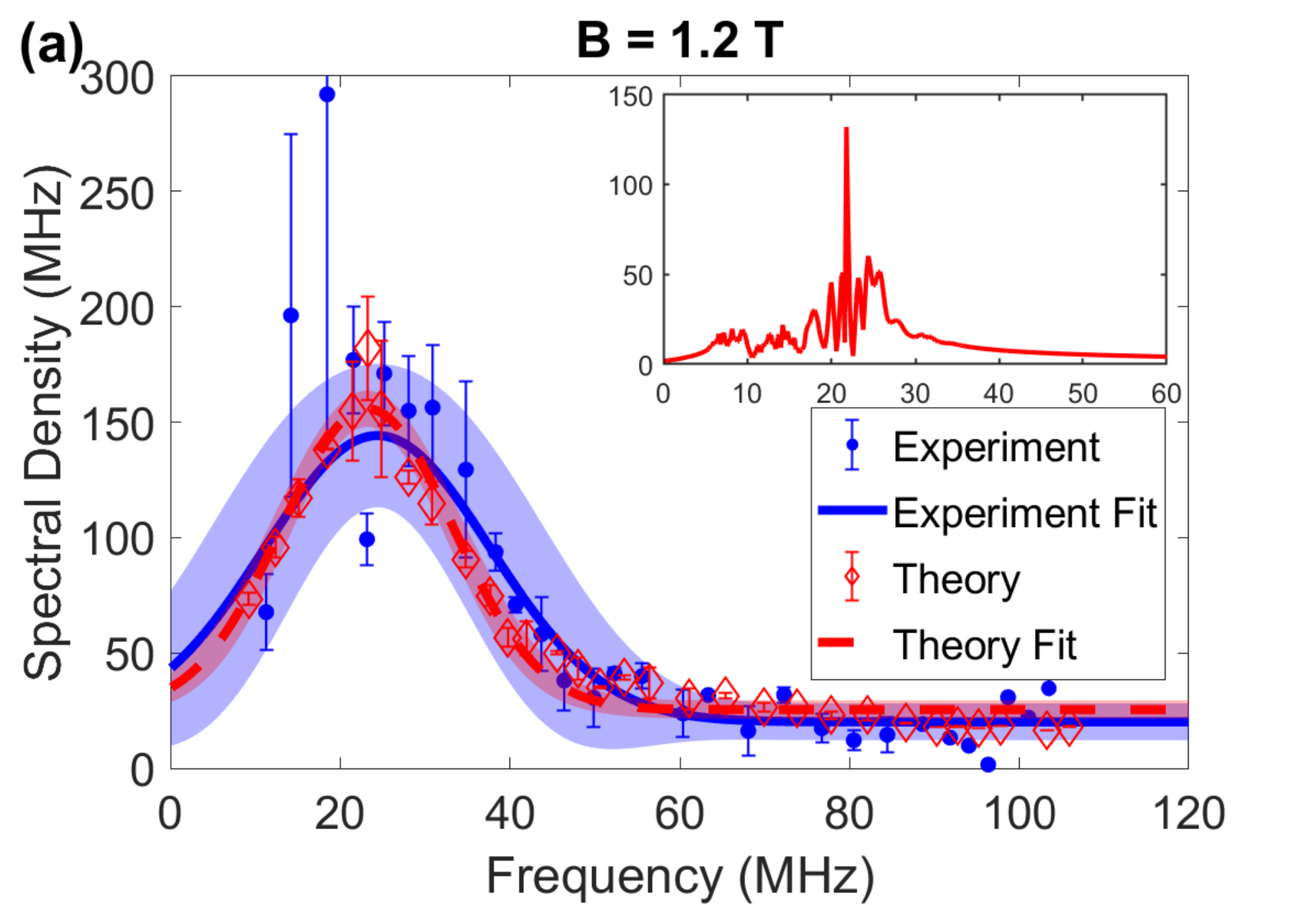}
	\includegraphics[width=0.32\textwidth]{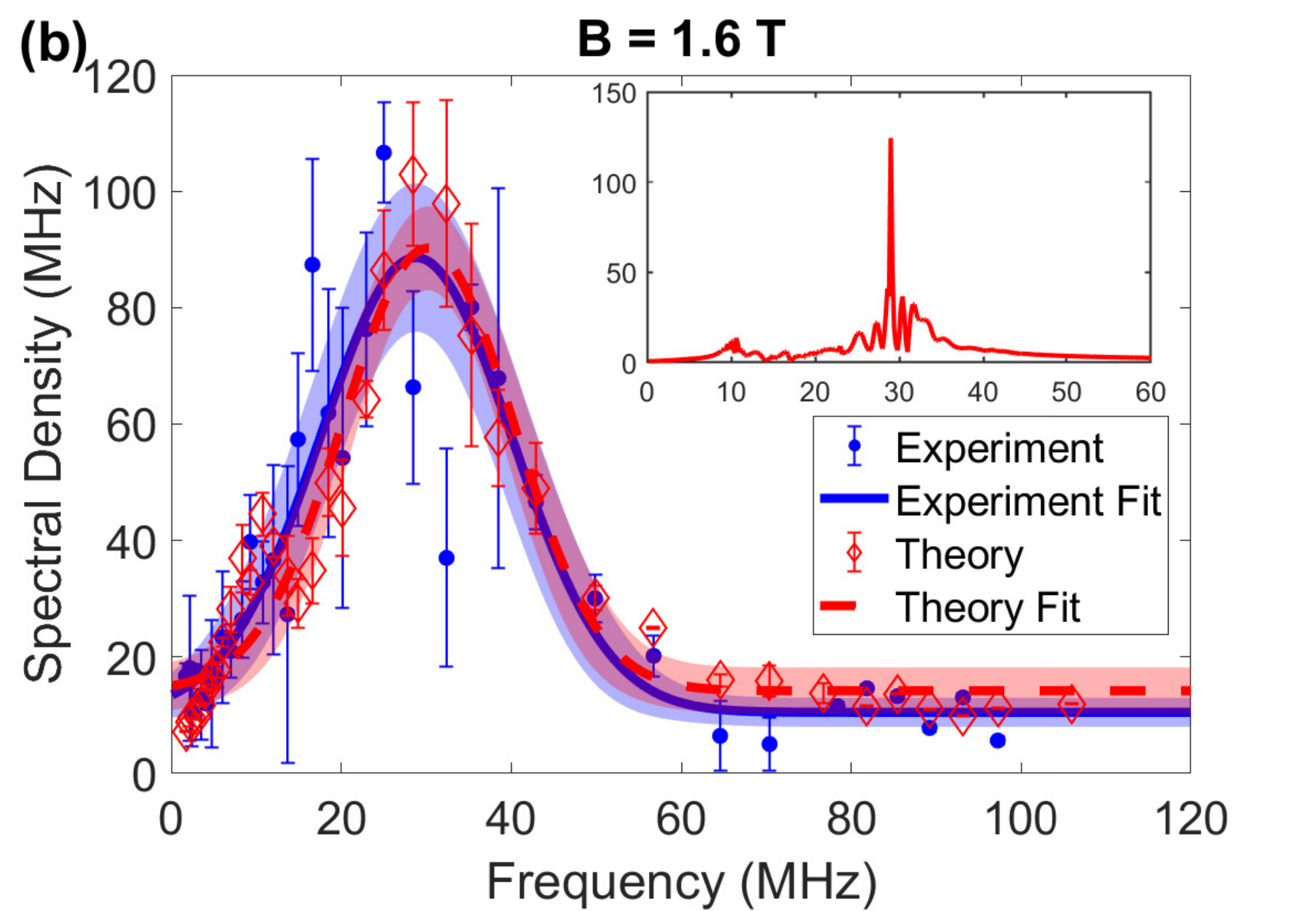}
	\includegraphics[width=0.32\textwidth]{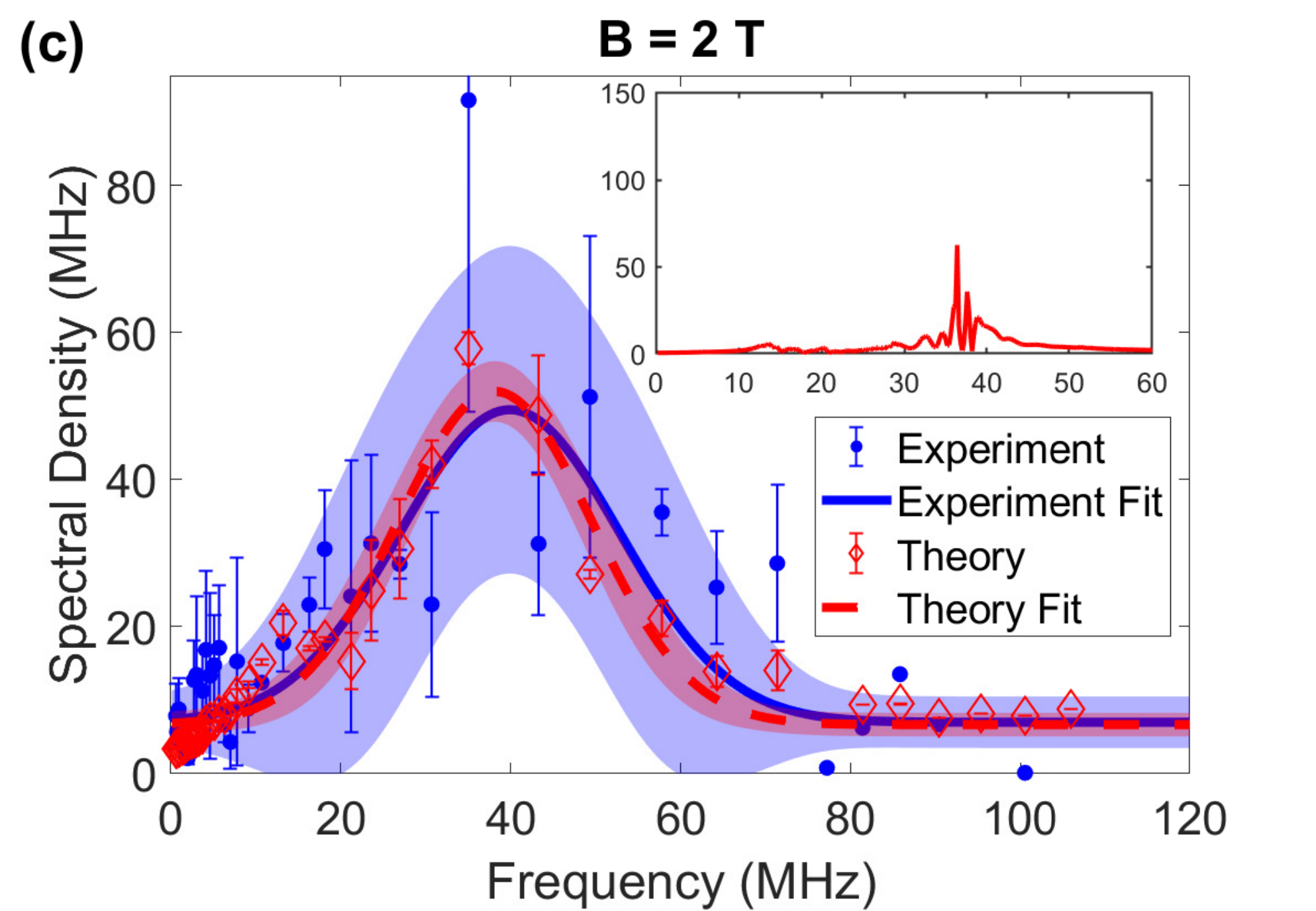}
	\includegraphics[width=0.32\textwidth]{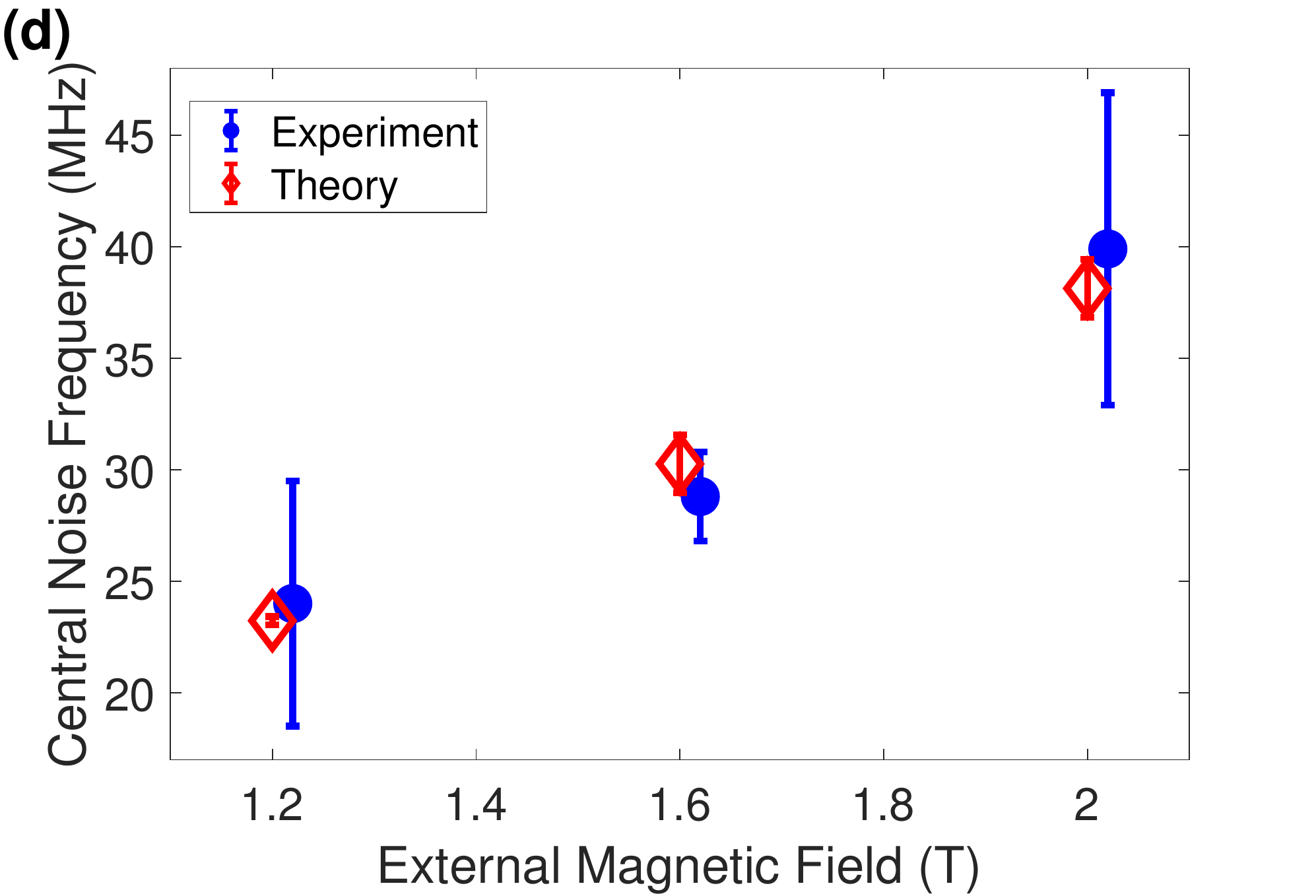}
	\includegraphics[width=0.32\textwidth]{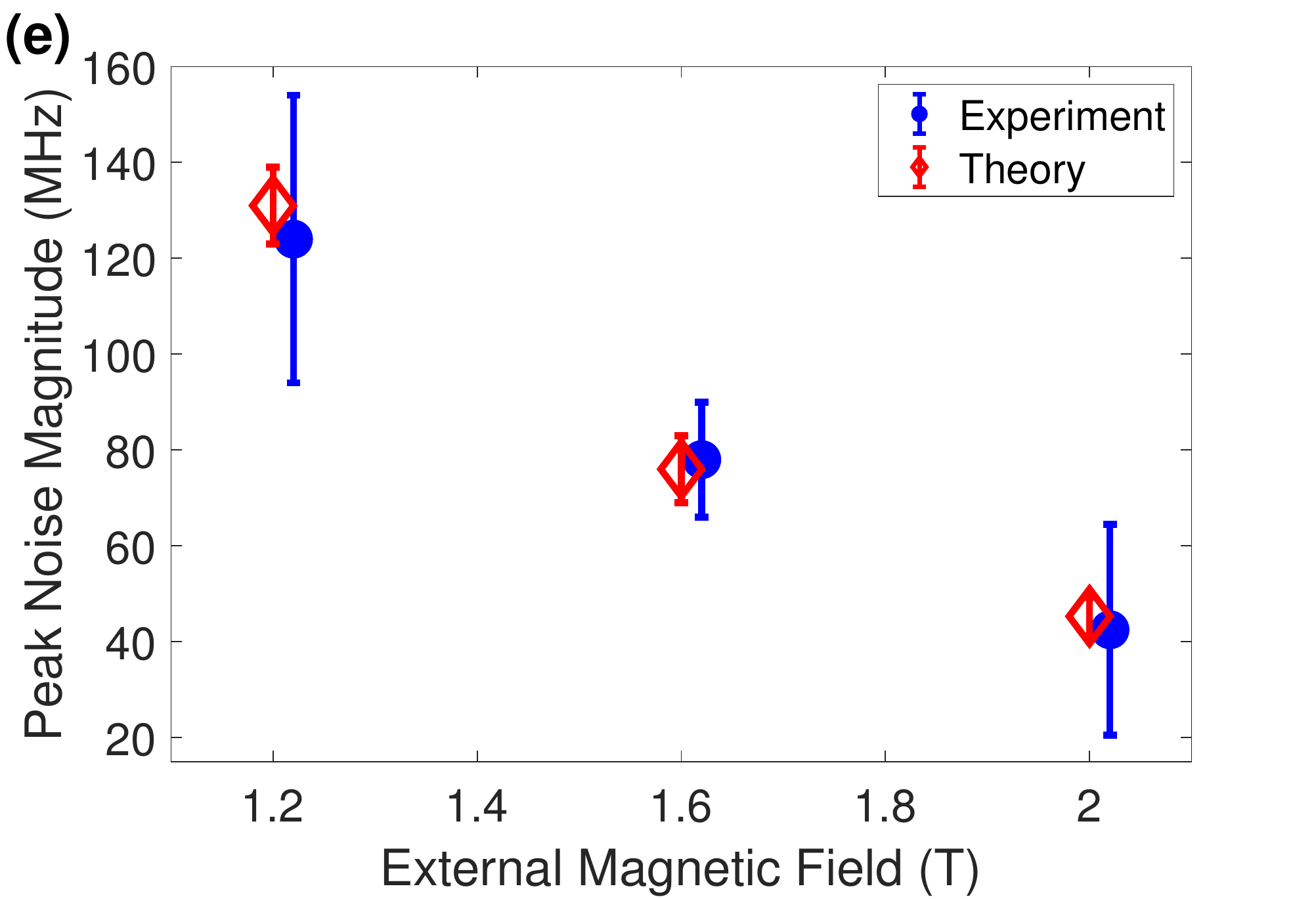}
\caption{
	(a)-(c) Noise spectroscopy of the nuclear spin ensemble interacting with a quantum dot, derived from spectral decomposition of CPMG experiments with $n = 1, 2, 4$ and $8$ pulses at varying magnetic fields of $B = 1.2$ T (a), $B = 1.6$ T (b), and $B = 2$ T (c). The blue dots represent experimental results, the solid blue lines represent the least-square fitting of these results to Gaussian functions, and the shaded blue area represents the uncertainties of the fits. The red diamonds are the result of applying the algorithm for noise spectroscopy on the ideal CPMG coherence functions generated from the theoretically simulated noise spectra, the  dashed red lines represent fitting of these results to Gaussian functions, and the shaded red area represents the uncertainties of the fits. The error bars in (a) -(c) originate from the (x2) binning of the data for the clarity of their presentation. Insets: The theoretically simulated noise spectra considering 40,000 nuclear spins, which were analyzed to produce the red diamond data in the main figures. (d) The central frequency and (e)        magnitude at the central frequency of the noise spectral density as a function of the magnetic field, extracted from the Gaussian fitting in (a)-(c), with error bars obtained from a least-square fitting algorithm.
}
	\label{fig:fig3}
\end{figure}

The extracted noise spectra verify a previously established theoretical model of the Overhauser field \cite{Bulutay2012,Stockill2016}.   Using this model, we simulate the noise spectra that represent the hyperfine coupling of the quantum dot spin to indium and arsenic nuclear spins experiencing quadrupolar coupling to strain fields (insets of Figure 3a-c), which exhibit sharp peaks that correspond to the different nuclear spin numbers. To compare the experimental results of noise spectroscopy with theory, we first use the theoretical spectra to calculate CPMG coherence functions with $n = 1, 2, 4$ and $8$ $\pi$-pulses under ideal conditions. We then apply the algorithm used to extract noise spectra from the experimentally obtained coherence functions on the simulated coherence functions (red diamonds in Figure 3a-c).
As the CPMG sequences probe the noise with spectrally broad filter functions that have high harmonics (Figure S3 of the Supporting Information) rather than with delta-like filter functions that probe the noise at single frequencies, the sharp peaks in the simulated spectra are broadened by the finite sampling resolution, and can be fitted to Gaussian functions similarly to the experimental results. The dashed red lines in Figure 3a-c that represent such Gaussian fits lie well within the uncertainties of the experimentally fitted results (shaded blue areas), thereby confirming the agreement between theory and experiment. Furthermore, the amplitudes of the simulated spectra (red diamonds in Figure 3e) consistently fit the experimentally extracted amplitudes and indicate that the quantum dot spin interacts with 40,000 nuclei (Section II of the Supporting Information), in agreement with common predictions \cite{Stockill2016}. 

The theoretical model of the quantum dot environment verified by all-optical noise spectroscopy can shed light on the coherent behavior of the quantum dot spin dynamics (e.g., the coherence functions in Figure 2). The model predicts two separate noise terms, perpendicular and parallel to the external magnetic field \cite{Bulutay2012,Stockill2016}. The noise component perpendicular to the external field, $S_\perp(\omega)$, monotonically decreases with the frequency and qualitatively fits the experimentally extracted noise floor (i.e., the baselines of the Gaussian fits in Figure 3a-c). This noise component dominates at low frequencies ($< 10$ MHz, Section II of the Supporting Information), thereby leading to the spin dynamics at long timescales ($> 100$ ns) depicted in Figure 2. Since the application rate of  $\pi$-pulses in our CPMG sequences is faster than the low frequencies of $S_\perp(\omega)$, increasing the number of pulses slows down the decay of the spin dynamics at long timescales. For example, under an external field of 2 T (Figure 2c), the decay of the spin dynamics at $T > 100$ ns is slower for $n = 4$ (dotted green line) than for $n = 1$ (solid blue line). In addition to the perpendicular noise term, the theoretical model predicts high frequency noise, $S_\parallel(\omega)$, which arises in parallel to the direction of the external field \cite{Bulutay2012,Stockill2016}. This parallel term is stronger than the perpendicular term and consists of peaks corresponding to the nuclear Larmor frequencies broadened by the environmental strain field (Section II of the Supporting Information). The broad spectral features of $S_\parallel(\omega)$ lead to the observed contrast drop \cite{deSousa2009} at the short timescales depicted in Figure 2. For the external magnetic field of $B = 1.2$ T, the large magnitude of $S_\parallel(\omega)$ leads to the complete loss of the quantum dot spin coherence under the  Hahn-echo sequence (dash-dotted black line in Figure 2a) at $T \approx 30$ ns, thus a second decay caused by  $S_\perp(\omega)$ is not observed for this field. The contrast drop caused by $S_\parallel(\omega)$ also quantifies the ability of the CPMG sequences to extend the quantum dot spin coherence time.  For example, the application rate of the $\pi$-pulses in our CPMG sequences is slower than the high frequency components of $S_\parallel(\omega)$ for $B = 2$ T. As a result, the decay of the spin dynamics at timescales shorter than 100 ns does not improve by increasing the number of pulses (Figure 2c).  By analyzing the obtained noise spectra, we learn that mitigating such high frequency noise to extend the quantum dot spin coherence time from the current state-of-the-art of a few microseconds  \cite{Stockill2016,Press2010,Bechtold2015} to beyond $10$ \SI{}{\micro\second} requires the application of dynamical decoupling sequences utilizing hundreds of $\pi$-pulses.
     
     However, here we are able to apply just eight $\pi$-pulses due to two mechanisms of spin relaxation. First, the natural spin relaxation of the quantum dot in our sample,  $T_1\approx 1$ \SI{}{\micro\second}, did not allow us to observe the expected spin dynamics under multi-pulse sequences beyond microsecond timescales. This relaxation time could be extended up to milliseconds by modifying the tunnel barrier of the sample \cite{Gillard2021}. Second, increasing the number of CPMG pulses in our measurements resulted in a dramatic contrast drop of the collected fluorescence as a function of $n$ \cite{Bodey2019} (Section V of the Supporting Information). This contrast drop is related to electron tunneling due to the increase of the laser power associated with the addition of the pulses. Such electron tunneling could be mitigated by coupling  the quantum dot to fabricated photonic structures that reduce the laser power required for spin rotation.  The reduction of the rotation laser power could enable the realization of multi-pulse sequences for prolonging spin coherence times and for the preservation of arbitrary spin states (e.g., XY8-based sequences \cite{Farfurnik2015}) for quantum information processing. Furthermore, the implementation of multi-pulse sequences with ultra high spectral resolutions (e.g., the "DYSCO" sequence \cite{Romach2019}) may enable the identification of individual Larmor frequencies associated with nuclear species (e.g., the peaks in the insets of Figure 3), as well as the ultrahigh resolution probing of external fluctuating magnetic fields with a single spin.

     To conclude, we introduce an all-optical approach for noise spectroscopy and implement it to study the environment of InAs quantum dots, for which the application of microwave control is challenging. The high Rabi frequencies and precise control capabilities of the Raman approach provide spectral bandwidths ($> 100$ MHz) that enable the identification of high frequency noise spectra. Leveraging the diffraction-limited spatial resolution provided by the optical fields could enable the measurement of noise correlations within neighboring regions of a given sample. 
     Performing all-optical noise spectroscopy while modifying  material properties (e.g., concentrations of dopants, layer thicknesses) could lead to the design of novel semiconducting heterostructures for quantum information processing in a variety of materials that host optically-active spins. The availability of optical coherent control in the solid-state community could enable such studies of spins in direct narrow bandgap semiconductors such as AlGaAs  \cite{Huber2017}, InP \cite{Benyoucef2013} and ZnSe \cite{Greilich2012}, as well as color center spins  with large spin-orbit couplings in wide bandgap semiconductors such as diamond \cite{Debroux2021}. In addition, noise spectroscopy of spins coupled to photonic structures could quantify the impact of the structures on spin coherence, thereby leading to the design of optimal photonic platforms for spin-photon interfaces. The experimentally extracted spectral densities, $S(\omega)$, can be plugged into equations of spin dynamics under the application of protocols for quantum sensing, communication, and information processing, thereby evaluating the potential of optically-active spin qubits for quantum technologies.

\subsection*{Supporting Information}
Experimental methods, theoretical modeling of the noise, optimization of experimental parameters, spectral decomposition algorithm, and studies of laser-induced tunneling.

\begin{acknowledgement}
We thank M. Atatüre, C. Le Gall, D. A. Gangloff, N. Bar-Gill, Y.    Romach, and D. S. Smirnov for fruitful discussions. This work has been supported by the   Physics Frontier Center at the Joint Quantum Institute, the National Science Foundation (Grants PHY1415485 and ECCS1508897), the U.S.-Israel Binational Science Foundation (Grant $\#$2018719), and the ARL Center for Distributed Quantum Information (Grant W911NF1520067). D. F.  acknowledges support by the Israel-U.S. Fulbright Postdoctoral Fellowship and the     Israel Council for Higher Education Quantum Science and Technology Scholarship. A.S.B. and S.G.C acknowledge support from the U.S Office of Naval Research. R.M.P. acknowledges support through an appointment to the Intelligence Community Postdoctoral Research Fellowship Program at the University of Maryland, administered by Oak Ridge Institute for    Science and Education through an interagency agreement between the U.S. Department of Energy and the Office of the Director of National                Intelligence. 
\end{acknowledgement}

	\bibliography{../../../../mytex/mybibliography}

\end{document}